\documentclass[aps]{revtex4}

\def\half {{1\over 2}}

\def\bea{\begin{eqnarray}}
\def\eea{\end{eqnarray}}
\def\ket#1{\vert #1 \rangle}

\def\sqr#1#2{{\vcenter{\vbox{\hrule height.#2pt
      \hbox{\vrule width.#2pt height#1pt \kern#1pt
         \vrule width.#2pt}
      \hrule height.#2pt}}}}

\begin{document}
\title{ Temperature of a Decoherent Oscillator with Strong Coupling }

\author{W. G. Unruh}
\affiliation{ CIAR Cosmology and Gravity Program\\
Dept. of Physics\\
University of B. C.\\
Vancouver, Canada V6T 1Z1\\
~
email: unruh@physics.ubc.ca}

~

~

\begin{abstract}
The temperature of an oscillator coupled to the vacuum state of a  heat bath
via ohmic coupling is non-zero, as measured by the reduced density
 matrix of the oscillator. This paper shows that the actual temperature, as measured by a thermometer is still zero (or in the thermal state of the bath, the temperature of the bath). The
decoherence temperature is due to "false-decoherence", with the heat bath
state being dragged along with the oscillator.

\end{abstract}

\maketitle

There are at least two ways of defining the temperature of a system.
 Assuming that the system is in equilibrium, one definition would be to
 look at the reduced density matrix of the system under consideration.
 If the system is in thermal equilibrium one would expect that reduced 
density matrix to have the form
\bea
\rho= N e^{-H\over T}
\eea
where $N$ is a normalisation factor such that $Tr(\rho)=0$, $H$ is an
 (effective) Hamiltonian for the system, and $T$ is the temperature of 
the system. 

An alternative way of defining the temperature is that one measures 
it with a thermometer. One (weakly) couples a system to the object 
under consideration, and looks at some physical feature of that 
system which depends on its temperature. 

One of the key features of a thermal density matrix is that the density matrix is not that of a pure state. It is a mixed state, in which there are no correlations between the various eigenstates of the Hamiltonian H. Another feature of 
the reduced density matrix is that it is ``instantaneous"-- one calculates 
the reduced density matrix at some instant in time. 

In a previous paper, I defined the concept of ``false
decoherence" \cite{falsedec}.  This is a situation in which the decoherence of the density matrix
of a system could disappear under certain circumstances. One can make certain measurements on the system which would show the state of the system to be that of a pure
state. In that case, the instantaneous measurement of the system was such that
at one time, the density matrix of the system was such that it was almost
completely decohered (for a two level system, was the identity matrix) while
at a later time the density matrix again looked like an almost pure state. 
 This situation differs from quantum recurrence,
 where the dynamics of the heat bath which cause the decoherence recur
 (as in spin echo). Instead the false decoherence occurs because the 
decoherence is essentially the result of correlations between the object and
 the heat bath which remain localized around the object, and depend on the state
of the object. Thus when one traced out the state over the heat bath, one
could find that that tracing could at one time produce a decoherent state for
the system itself, while at a later time, that tracing would leave the system
in a pure state. 

A hand-waving example of this is to regard and electron taking two different
paths through an electron microscope. If one traces out over the degrees of
freedom of the electromagnetic field, one finds that the Coulomb filed of the
electron decoheres the state of the electron. The overlap integral of the
Coulomb fields of an electron at distant points in space goes to zero, and the
state of the electron decoheres if widely separated. But if the electrons are
then re-guided back to a common point, the electron having travelled along
different paths will again interfere, since the Coulomb field of the electron
located at nearby positions will again overlap, and the decoherence which
arose out of that lack of overlap will again disappear. (Note that this
differs from the situation in which the electron radiates differently along
its different possible paths. In that case the electromagnetic field will
cause genuine, long lasting, loss of coherence).

In this paper I will look at another instance of such false decoherence.
 In this case the object of interest will be a simple Harmonic oscillator 
coupled to a one dimensional scalar field as a heat bath. The scalar
 field will be taken to be a field in a vacuum state.
 Because of the strong coupling of the field to the oscillator, the reduced density
 matrix of the oscillator will not be in its ground state, but rather in a
 thermal state with some effective Hamiltonian defining that thermal state.
I will model the thermometer as another harmonic oscillator very weakly coupled
to the first oscillator, to minimize the impact of that second oscillator on
the first. In all cases I will wait long enough that any transients will have
died away. I will demonstrate that just as in the cases discussed
previously, this system also shows an aspect of false decoherence. The
density matrix of the first oscillator will be shown to be in a thermal state 
with a possibly large temperature, and a large entropy. Nevertheless, the 
thermometer, i.e., the second oscillator which couples only weakly to the first oscillator,
will be shown to be in its ground state, a completely coherent state. It 
measures the temperature of a system with a thermal density matrix as a completely
coherent state. 

Note that a one dimensional massless scalar field radiates extremely easily.
There are no phase space barriers to radiation (the volume of phase space is
the same at all energies). Thus one would expect any system coupled to a
massless two-dimensional scalar field to decoher very easily. Nevertheless,
we shall find that, while the coupling of the oscillator to the scalar field
in its vacuum state 
indeed does decoher the oscillator, there are measurements one can make on
the oscillator which show it to be in a coherent state. In particular the
thermometer attached to the oscillator shows it to have zero temperature.

\section {Strongly Coupled Oscillator}

Let us consider an oscillator, with dynamic degree of freedom $q$ coupled to a
heat bath represented by a scalar one dimensional field $\phi$ and the oscillator located at  position
$x=0$ in the space of the scalar field. The total action is given by
\bea
I=\half (\dot q^2-\Omega^2 q^2) +\half \int (\dot\phi^2- \phi'^2 )+2\epsilon
q\dot \phi(t,0)
\eea
where $\dot{}$ denotes the time derivative and ${}'$ the spatial ($x$)
derivative.
The equation of motion for the field is
\bea
\ddot\phi-\phi'' =-\epsilon\dot q \delta(x)
\eea
which has the solution
\bea
\phi=\phi_0(t,x) -{\epsilon\over 2}q(t-|x|)
\eea
where $\phi_0$ is the ``incoming" free field.

Substituting this into the equation of motion for $q$ gives
\bea
\ddot q +{\epsilon^2\over 2}\dot q + \Omega^2 q =\epsilon\dot\phi_0(t,0)
\eea
This model thus has the  standard Ohmic damping  for the oscillator.  Writing $\phi_0$ in
terms of the right and left movers $\phi_0(t,x)= \phi_+(t-x) +\phi_-(t+x)$, we
have (with $q(t)= \int q(\omega) {e^{i\omega t}\over \sqrt{2\pi}}$)
\bea
q(\omega)= {i\epsilon\omega \over -\omega^2+i{\epsilon^2\over 2}\omega
+\Omega^2} \left(\phi_+(\omega)+\phi_-(\omega)\right)
\eea

The field operators $\phi_{\pm}(\omega)=\int\phi_\pm (t)e^{-i\omega t}
{dt\over\sqrt{2\pi}}$ are creation/annihilation operators
such that
\bea
[\phi_{\alpha}(\omega), \phi_{\beta}(\omega')]=
{1\over 2\omega}\delta_{\alpha\beta}\delta(\omega+\omega')
\eea
and the vacuum state for the incoming scalar field is given by
\bea
\phi_\pm (-|\omega|)\ket{0}=0
\eea
which gives
\bea
\langle \phi_\alpha(\omega) \phi_\beta(\omega')\rangle=-{1\over 2\omega} \delta_{\alpha\beta}
\delta(\omega+\omega')\Theta(-\omega)
\eea
where $\alpha,\beta$ have values $\pm$, and $\Theta$ is the Heaviside step
function. 

We are interested in the correlations $<q^2>$, $<pq+qp>$, and $<p^2>$ in the
limit as $t$ gets very large so that any initial conditions of the oscillator
have died out. 

We have
\bea
<q^2>&=& \int_0^\infty {\epsilon^2\over 2\pi}{ \omega^2\over  (\omega^2-\Omega^2)^2
+({\epsilon^2\omega\over 2})^2}\\
<qp+pq>&=& 0\\
<p^2>&=& \int_0^\infty {\epsilon^2\over 2\pi}{ \omega^3\over (\omega^2-\Omega^2)^2
+({\epsilon^2\omega\over 2})^2}d\omega
\eea

Note that for $<p^2>$, the integral diverges logarithmically at the upper end,
and requires a  cutoff to keep it finite. This cutoff could be
introduced in the interaction between the oscillator and the field (e.g., by replacing
 the delta function interaction between the field and the oscillator
 with some sharply peaked function $\int\dot\phi(t,x)q(t)dx$, or it could be introduced
in the interaction between the oscillator and any measuring apparatus 
which attempts to measure its momentum: if that
measurement had some finite duration, that would also provide an effective
cutoff. I will just take a simple cutoff frequency $\hat\Omega$ to the
integral, with $\epsilon^2<<\hat\Omega>>\Omega$.

We can also write the two time correlation function for $q$ as
\bea
<q(t)q(t')+q(t')q(t)>= \int_0^\infty {2
\epsilon^2\over 2\pi}{ \omega\over 
(\omega^2-\Omega^2)^2
+({\epsilon^2\omega\over 2})^2} \cos(\omega(t-t'))d\omega
\eea
in which case
\bea
<qp+pq>= {d<q(t)q(t')+q(t')q(t)>\over dt} \vert_{t'=t}=0
\eea
because the correlation function is symmetric in $t-t'$ and its derivative is
antisymmetric. Furthermore,
\bea
2<p^2>&=& {d^2<q(t)q(t')+q(t')q(t)>\over dtdt'}\vert_{t'=t}= <-{d^2q\over dt^2}
q(t')-q(t'){d^2q\over dt^2}>\vert_{t'=t}\\
&=&{\epsilon^2\over 2}<pq+qp>+\Omega^2 <q^2> +\epsilon<\phi_0 q+q\phi_0>
\eea
The last term is proportional to $\epsilon^2$ and is the term which gives the 
logarithmic divergence in $<p^2>$.

The expression $<p^2><q^2>-{1\over 4}<pq+qp>^2$ is invariant under any 
linear symplectic transformations of $p$ and $q$, (i.e., a transformation which
leaves the form $p\dot q$ invariant up to a complete time derivative).
These can be made up of
successive applications of the transformations of the form, $\tilde p=\alpha p,~~\tilde
q={1\over \alpha} q$; $\tilde p= p+\alpha q,~~\tilde q=q$; or $\tilde
p=p,~~\tilde q=q+\alpha p$.

The reduced density matrix for the oscillator after integrating over the
fields will be a quadratic Gaussian form, so that 
\bea
\rho =Ne^{-\tilde H/T}
\eea
for some quadratic form $\tilde H$ and some temperature $T$. $N$ is a
normalisation constant, so that $Tr(\rho)=1$. (If $T\rightarrow
0$, it will be a Gaussian pure state). 
One can always make a symplectic transformation so that any quadratic form of
the $p$ and $q$ has the form 
\bea\tilde H={1\over 2} \Lambda (\tilde p^2+\tilde q^2)
\eea
 for
which a 'thermal state' $\rho= Ne^{-\tilde H/T}$ has the properties that 
\bea
<\tilde p^2>=<\tilde q^2>={1\over 2}\coth\left({\Lambda\over 2T}\right)\\
<\tilde p\tilde q+\tilde q\tilde p>=0
\eea
Thus, \bea
<p^2><q^2>-{1\over 4}<pq+qp>^2 ={1\over 4}\coth^2\left({\Lambda\over 2T}\right)
\eea

So, unless $<p^2><q^2>-{1\over 4}<pq+qp>^2={1\over 4}$, the reduced density
matrix of the oscillator is a thermal state with a non-zero entropy and
temperature. Only in
the limit as $\epsilon$ goes to 0 is this effective temperature equal to 0.

Evaluating $<q^2>$ and ignoring terms which go as inversely as the cutoff,  we
have for $\epsilon^2<4\Omega$
\bea
<q^2>&=&\int {\epsilon^2\over \pi}{\omega\over (\omega^2-\Omega^2)^2 +\omega^2{\epsilon^4\over
4}} d\omega\\
&=&{\epsilon^2\over 2\pi} \int {dy\over (y-\Omega^2)^2+y{\epsilon^4\over 4}  }\\
&=& {\epsilon^2\over 2\pi}{1\over \sqrt{\epsilon^4\Omega^2 -\epsilon^8/16}}
\left[2\pi -2{\rm atan}\left({2\epsilon^2\sqrt{16\Omega^2-\epsilon^4}\over
8\Omega^2-\epsilon^4}\right) \right]
\eea
For small $\epsilon$ we get 
\bea
<q^2>= {1\over 2\Omega} +O(\epsilon^4)
\eea

If $\epsilon^2>4\Omega$, we have
\bea
<q^2>= {1\over\sqrt{\epsilon^2/16-\Omega^2}}\ln\left( {\epsilon^4-16\Omega^2
+2\epsilon^2\sqrt{\epsilon^4-16\Omega^2}\over \epsilon^4-16\Omega^2
-2\epsilon^2\sqrt{\epsilon^4-16\Omega^2} }\right)
\eea

For $<p^2>$ we have
\bea
<p^2>=(2\Omega^2-\epsilon^4/4) <q^2> +{2\epsilon^2\over \pi}\ln(\left( {\hat\Omega\over \Omega}\right)
\eea
Similarly 
\bea
<p^2>= \Omega^2 <q^2> +{\omega^2\epsilon^2 \over
(\omega^2-\Omega^2)+i\omega\epsilon^2/2}....
\eea

\section{Temperature via thermometry}

The density matrix of the oscillator looks very much like a thermal state notwithstanding its Hamiltonian which is not exactly the same as the original uncoupled
Hamiltonian. But is this really a temperature? To answer this question we can
try to couple a thermometer to the oscillator. In this case my thermometer
will be another harmonic oscillator weakly coupled to the original one. We
know that in the limit as the coupling strength goes to 0, the temperature of
the oscillator is the temperature of the system it is coupled to. 

Thus we take the Lagrangian density as 
\bea
L=\int( {1\over 2} (\dot \phi ^2-\phi'^2 ) +\epsilon \dot\phi q \delta (x) )dx
+{1\over 2}(\dot q^2-\Omega^2q^2) +\mu qz +{1\over 2}(\dot z^2+\Lambda^2 z^2)
\eea
The solution for the thermometer is
\bea
z(\omega)= -\mu \epsilon {i\omega\over (-\omega^2+\Lambda^2)(-\omega^2+i\omega
\epsilon^2/2 +\Omega^2) -\mu^2} (\phi_+(\omega)+\phi_-(\omega))
\eea
and 
\bea
<z(-\omega)z(\omega)> = {2\over 2\pi} \mu^2\epsilon^2 {\omega\over
((-\omega^2+\Lambda^2)(-\omega^2+\Omega^2)-\mu^2)^2+(-\omega^2+\Lambda^2)^2\omega^2\epsilon^4/4}
\eea
We want the limit of this as $\mu\rightarrow 0$. In that limit, the only terms
that will survive in the integral to find $<z^2>=\int_0^\infty <z(-\omega)z(\omega)>d\omega$
are those due to the poles at
\bea
\omega \approx \Lambda +{ \mu^2\over 2\Lambda (\Lambda^2-\Omega^2 \pm
i{\epsilon^2\over 2})}
\eea
 which in the limit as $\mu\rightarrow 0$ give
\bea
<z^2>= {1\over 2{\Lambda}}+O(\mu^2)
\eea
Similarly, the momentum is 
\bea
<p_z^2>= {{\Lambda}\over 2}+O(\mu^2)
\eea
and $<p_z z+z p_z>=0$. In the limit as $\mu\rightarrow 0$ and at long
enough times such that the thermometer has come into equilibrium,  the thermometer is in its ground state, with a
temperature of 0 (to order $\mu^2$). 

The $q$ oscillator is, at all times,  in a time independent mixed state. The thermometer
interacts only with that system in a mixed state, and one would thus expect it
also to be in a mixed state. However, the termometer at long times ends up in
a pure state, with a temperature of 0.

Of course, if one regards the thermometer as interacting with the complete
system, the $q$ oscillator and its heat bath, then, after a long time one
expects that system to be in its ground state, with temperature 0, and thus
the themometer would also be expected to read 0, as it does.

This is another example of "false decoherence". As in that previous work, the
apparent decoherence of the spin system did not preclude interference between
the "decohered" states of the spin, as long as the process revealing that
interference occurred slowly enough, here again, the thermometer interacting
slowly enough with the $q$ oscillator displays interference between the various
occupied energy levels of that oscillator, leaving the thermometer in its
ground state. 

This also suggests that just as in the spin case, the "true" states of the
oscillator in the adiabatic limit are not the energy eigenstates, but rather
are correlated states between the oscillator and the bath represented by the
field $\phi$. The surprizing feature here is that the bath is one with no
energy gap. It is a massless field, which can carry off energy, and  entropy
or coherence at as low a frequency and energy as desired. There is nothing
that is adiabatic as far as the the $\phi$ field is concerned.

\section{Correlations}

In order to better understand these correlations between the oscillator and
field, let us calculate them to see what they are. The field $\phi$,
its conjugate momentum 
\bea
\pi= \dot \phi(t,x) +\epsilon q(t) \delta(x)
\eea
 commute with both  the position $q(t)$ and momentum $p(t)=\dot q$ of the
oscillator. 
Thus I will be interested in the correlation functions 
 $q(t)\phi(t',x)$ from which the equal time commutation
relations can be calculated. In order to do so, we look at the correlators of
the Fourier transforms
\bea
&\langle& q(\omega)\phi(\omega',x)\rangle \\
&&=\delta(\omega+\omega') {\left[
i\epsilon\omega {\langle (\phi_+(\omega,0) +\phi_-(\omega,0))
(\phi^+(-\omega,x)+\phi_-(-\omega,x))
+i\gamma \omega (\phi_+(-\omega,0) +\phi_-(-\omega,0))e^{i\omega|x|}\rangle\over 
(\omega^2-\Omega^2)^2+\gamma^2\omega^2}\right]}\\
&&=i{\epsilon\over 2\pi}\delta(\omega+\omega')\Theta(\omega) \left[{e^{i\omega
|x|}\over (-\omega^2+\Omega^2-i{\epsilon^2\over 2}\omega)} + {e^{-i\omega |x|}
\over -\omega^2+\Omega^2+i{\epsilon^2\over 2}\omega)}\right]
\eea
and
\bea
\langle \phi(\omega',x)q(\omega)\rangle= -\langle
q(-\omega)\phi(-\omega',x)\rangle
\eea

This gives us that the commutator of $q$ and $\phi$ at equal times is
\bea
\langle q(t)\phi(t,x)-\phi(t,x)q(t)\rangle= 2i{\epsilon\over 2 \pi}
\int_{-\infty}^\infty  \left[ {e^{i\omega|x|}\over
(-\omega^2+\Omega^2-i{\epsilon^2\over 2}\omega)} 
\right] d\omega =0
\eea
as expected, and the correlation function is
\bea
\langle q(t)\phi(t,x)+\phi(t,x)q(t)\rangle=4i{\epsilon\over 2\pi}\int_0^\infty
\left[{e^{i\omega|x|}\over
(-\omega^2+\Omega^2-i{\epsilon^2\over 2}\omega)} 
\right] d\omega 
\eea
which can be written in terms of Exponential integrals. This correlation
function falls off as $e^{-{\epsilon^2\over 2}|x|}$. That is, the correlation
function falls off exponentially with distance from the oscillator, as one
would expect from the exponential decay of the oscillator. 

\section{Conclusion}

The result we found was that although the density matrix of an oscillator
strongly coupled to a scalar field leaves the oscillator in a state which is
that of a thermal density matrix, the attempt to measure the temperature of
that oscillator with a weakly coupled thermometer (in this case another
oscillator weakly coupled to the first) produces a value of very near zero.
This is surprizing since the thermometer couples only to the first oscillator,
which is in a mixed state. One would expect an object coupled to something in
a mixed state to also be in a mixed state. However, that mixture comes about
because of the coupling of the oscillator to another external heat bath which
is in its ground state. If the thermometer were to instantaneously measure
some aspect of the state of the oscillator it would find that oscillator to be
a mixed state. All measurements made on it rapidly would produce results
indistinguishable from the oscillator being in a thermal state with non-zero
temperature. However, if the measurements are made slowly enough (over a time
period long compared with the decay time of the oscillator), it is the heat
bath and not the oscillator that dominate the dynamics, and measurement over
that longer time period would find the oscillator to be in a pure state on
average. 

This is another example of false decoherence. While the oscillator looks to be
decoherent over short time intervals, if one waits a long time, makes
measurements averaged over a long time, that oscillator is in a pure state,
the ground state. 

This illustrates the dangers of of using some simple measure of coherence or
decoherence to make physical predictions. Decoherence is not some attribute
that a body has, and thereafter retains. It is a property that depends both on
its coupling to the external word, and on the time scale over which
measurements are made. 

Note that this result is not one about the ease of radiation emitted by a
body. In the previous paper, the heat bath was a massive scalar field, and  the energy of the oscillator was less than the
mass of the scalar field. This meant that the system could not emit radiation
at its natural frequency into the heat bath. It could still distort the field,
so that the states of the field created by the different states of the system
had a small overlap, and thus caused decoherence of the system when one traced
out over the states of the field. However in the  case discussed here, because of its coupling to a massless two dimensional field, the
oscillator can radiate into that scalar field easily on any time scale. There
is no inherent time scale associated with the scalar field. Furthermore, in the
case of our thermometer, its dynamics is at its natural frequency, and at that
frequency the scalar field has no problem supporting outgoing radiation, either energetically or in terms of phase space, in carrying away energy. Furthermore there is no question of
Poincare recurrence or any other kind of recurrence. Once the scalar field is
excited, all excitations are carried away at a fixed velocity, never to
return. It is an infinite heat bath which accepts, and never returns, anything
which is consigned to its depths. Nevertheless, the thermometer measures the
oscillator to have zero temperature, to be in its ground state, if the
coupling between the thermometer and the oscillator is small enough. 

\begin{acknowledgments}
I would like to thank R. Parentani for asking the question (what temperature
would a thermometer measure if coupled to a system strongly coupled to a zero
temperature heat bath) at a meeting in the Peyresq School of Cosmology 
 which this paper answers. I would also thank the Canadian Institute for
Advanced Research (CIfAR) and the Natural Sciences and Engineering research
council of Canada for their support of my research while this work was carried
out. Some of the writing of this paper also occured while visiting the
Perimeter Institute as a Distinguished Research Chair. 
\end{acknowledgments}

\end{document}